\documentclass[superscriptaddress,twocolumn,showpacs,preprintnumbers,prb,groupedaddress]{revtex4}

\usepackage{graphicx}% Include figure files
\usepackage{subfigure}
\usepackage{amsmath}
\usepackage{subfigure}
\usepackage{booktabs}
\usepackage{slashbox}

\begin{document}
\date{\today}

\title{Density Matrix Renormalization Group Study of a Quantum Impurity Model with Landau-Zener Time-Dependent Hamiltonian}

\author{Cheng Guo}
\affiliation{Physics Department, Arnold Sommerfeld Center for Theoretical Physics, and Center for NanoScience, Ludwig-Maximilians-Universit{\"a}t M{\"u}nchen, D-80333 M{\"u}nchen, Germany}
\affiliation{Institute of Theoretical Physics, Chinese Academy of Sciences, P.O. Box 2735, Beijing 100080, China}
\author{Andreas Weichselbaum}
\author{Stefan Kehrein}
\affiliation{Physics Department, Arnold Sommerfeld Center for Theoretical Physics, and Center for NanoScience, Ludwig-Maximilians-Universit{\"a}t M{\"u}nchen, D-80333 M{\"u}nchen, Germany}
\author{Tao Xiang}
\affiliation{Institute of Physics, Chinese Academy of Sciences, P.O. Box 603, Beijing 100080, China}
\affiliation{Institute of Theoretical Physics, Chinese Academy of Sciences, P.O. Box 2735, Beijing 100080, China}
\author{Jan von Delft}
\affiliation{Physics Department, Arnold Sommerfeld Center for Theoretical Physics, and Center for NanoScience, Ludwig-Maximilians-Universit{\"a}t M{\"u}nchen, D-80333 M{\"u}nchen, Germany}

\begin{abstract}
We use the adaptive time-dependent density matrix renormalization group method (t-DMRG) to study the nonequilibrium dynamics of a benchmark quantum impurity system which has a time-dependent Hamiltonian. This model is a resonant-level model, obtained by a mapping from a certain ohmic spin-boson model describing the dissipative Landau-Zener transition. We map the resonant-level model onto a Wilson chain, then calculate the time-dependent occupation $n_d(t)$ of the resonant level . We compare t-DMRG results with exact results at zero temperature and find very good agreement. We also give a physical interpretation of the numerical results. 
\end{abstract}

\pacs{03.65.Yz, 74.50.+r, 33.80.Be, 73.21.La}

\maketitle

\section{Introduction}

Quantum impurity models, describing a discrete degree of freedom coupled to a continuous bath of excitations, arise in many different contexts in condensed matter physics. In particular, they are relevant for the description of transport through quantum dots and of qubits coupled to a dissipative environment \cite{Nakamura99, Chiorescu03}. In recent years, there has been increasing interest in studying the real-time dynamics of such models for Hamiltonians $H(t)$ that are explicitly time-dependent, as relevant, for example, to describe external manipulations being performed on a qubit. It is thus important to develop reliable numerical tools that are able to deal with such problems under very general conditions.

The most widely used numerical method to study quantum impurity systems is Wilson's numerical renormalization group (NRG) \cite{WilsonNRG}. With the recently proposed time-dependent NRG (TD-NRG) \cite{tdnrg} one can now calculate certain class of time-dependent problems where a sudden perturbation is applied to the impurity at time $t=0$. TD-NRG may very well be accurate for arbitrary long time. However, up to now, TD-NRG is not capable of dealing with a Hamiltonian $H(t)$ with a  time-dependence more general than a single abrupt change in model parameters at $t=0$. We will show in this paper that the adaptive time-dependent density matrix renormalization group method (t-DMRG)  is a promising candidate for treating a general time-dependent Hamiltonian $H(t)$.

The density matrix renormalization group method (DMRG) is traditionally a numerical method to study the low lying states of one-dimensional quantum systems \cite{white93}.  The recent extension of this method, the adaptive time-dependent DMRG (t-DMRG) \cite{daley04,tdmrg}, can simulate real-time dynamics of one-dimensional models with time-dependent Hamiltonians as well. t-DMRG has already been used to study problems involving real-time dynamics of one-dimensional quantum systems, for example the far-from-equilibrium states in spin-1/2 chains \cite{Gobert05}, dynamics of ultracold bosons in an optical lattice \cite{Kollathpra05,Kollath06}, transport through quantum dots \cite{transportQDs06}, dynamics of quantum phase transition \cite{Pellegrini}, and demonstration of spin charge separation \cite{DMRGspincharge}. These works showed that t-DMRG is a versatile and powerful method to study the real-time dynamics of one-dimensional quantum systems.

The underlying mathematical structures of DMRG and NRG are similar in the matrix product state representation language \cite{mps_nrg}. Indeed, once a quantum impurity model has been transformed into the form of a Wilson chain model, it can be treated by DMRG instead of NRG \cite{mps_nrg, Saberi08, nishimoto_density-matrix_2004, da_silva_transport_2008}. This possibility opens the door toward studying time-dependent quantum impurity models using t-DMRG. In this paper, we take a first step in this direction by using t-DMRG to study a simple, exactly solvable quantum impurity model whose Hamiltonian is a function of time. This model allows us to benchmark the performance of t-DMRG by comparing its results to those of the exact solution.

\section{The Model and DMRG Method}
We study the resonant-level model with a time-dependent potential applied to the level. The Hamiltonian is 
\begin{equation}
\hat{H}(t)=\epsilon_d(t)d^\dagger d+\sum_k \epsilon_k c^\dagger_k c_k+V \sum_k (d^\dagger c_k+c_k^\dagger d).
\label{Horiginal}
\end{equation}
$d^\dagger$  creates a spinless fermion on the level (impurity) and $c^\dagger_k$ creates a spinless fermion with momentum $k$ in a conduction band whose density of states is constant between $-D$ and $D$ and zero otherwise, with Fermi energy set equal to 0. The energy of the local band is swept linearly with time, $\epsilon_d(t)=D v t$, where $v$ is the sweeping rate in units of the half band width $D$.  This model is equivalent to the dissipative Landau-Zener model with a Ohmic boson bath whose spectral function is $J(\omega)=2 \pi \alpha \omega$, for $\omega \ll \omega_c$, where $\omega_c$ is the high energy cutoff  \cite{leggett87}, and the dimensionless strength of dissipation parameter $\alpha$ is henceforth set equal to $\frac{1}{2}$. When $\alpha$ is close but not equal  to $\frac{1}{2}$, the Hamiltonian (1) contains an additional interaction term proportional to $U (d^\dagger d - \frac{1}{2})(\sum_{k,k'}c_k^\dagger c_{k'} - \frac{1}{2})$ \cite{Guinea85}, but this case will not be considered here.

At time $t_0 \to -\infty$ the local level contains a spinless fermion and the band is half filled. Then, we lift the energy of the level linearly with time. As the level approaches the band, the probability that the fermion jumps to and from the band will increase, and decrease after the level has passed the band. In this paper we study this problem in detail. In particular, we are interested in the expectation value of the occupation number on the level $n_d(t)$ at time $t$.

Before using t-DMRG to solve this problem, we need to transform the Hamiltonian to a DMRG-friendly form. This can be realized by using a standard Wilson mapping (originally invented in the context of NRG), which include two steps: logarithmic discretization of the band and converting the Hamiltonian to a hopping form \cite{nrg80,NRGreview}. Here, we just give the final result: The Hamiltonian~(\ref{Horiginal}) is mapped to a semi-infinite Wilson chain
\begin{eqnarray}
\label{Hwillsion}
 &&\hat{H}(t)=\epsilon_d(t) d^\dagger d + (\frac{2 \Gamma D}{\pi})^{\frac{1}{2}}(f^\dagger_0 d+d^\dagger f_0) \nonumber \\
&&+{D \over 2}(1+\Lambda^{-1})\sum_{n=0}^\infty \Lambda^{-\frac{n}{2}} \xi_n(f^\dagger_nf_{n+1}+f^\dagger_{n+1}f_n),
\end{eqnarray}
where $\xi_n=(1-\Lambda^{-n-1})(1-\Lambda^{-2n-1})^{-\frac{1}{2}}(1-\Lambda^{-2n-3})^{-\frac{1}{2}}$. 
$ \Gamma\equiv\pi\rho V^2$ is the hybridization parameter,  and $\rho$ is the density of states at the Fermi level. $\Lambda>1$ is a logarithmic discretization parameter, which means we divide the band into discrete energy intervals determined by $\pm\Lambda^{-1}, \pm\Lambda^{-2}, \pm\Lambda^{-3}, \cdots$. In the limit $\Lambda \to 1$, the discretized spectrum becomes dense throughout the band. The hopping factors in Hamiltonian~(\ref{Hwillsion}) decrease exponentially, so it is sufficient to keep the first $L$ sites to achieve an energy resolution of $\Lambda^{-L/2}$. 

The dimensionless parameter $r \equiv 2 \Gamma /v$ can be used to define three typical regimes of this problem. They are:
\begin{itemize}
 \item \textit{Fast sweep}: $r \ll 1$
 \item \textit{Intermediate sweep}: $r \simeq1$
 \item \textit{Slow sweep}: $r \gg 1$
\end{itemize}
We will examine the performance of DMRG in all these regimes.

The Wilson-chain form of Hamiltonian~(\ref{Hwillsion}) can now be treated using DMRG. We first use infinite and finite DMRG \cite{white93} to calculate the ground state of the initial Hamiltonian $\hat{H}(t_0)$ at $t_0$. This ground state is a very good approximation to the true initial state in the ideal case in which the level would start from $t_0 \to - \infty$ as long as $\epsilon_d(t_0) \ll -|\Gamma|$. In the fast and intermediate sweep regimes, we can choose $t_0$ so that the $\epsilon_d(t_0)=D v t_0$ is far below the Fermi surface to satisfy $\epsilon_d(t_0) \ll -|\Gamma|$. In slow sweep regime we can do the same if we use a very large $|t_0|$. However, a more efficient way we adopt is to use a moderate $t_0$, but set $\epsilon_d(t_0)$ as a very low value (e.g. $-10000D$). After we get the starting state we apply the evolution operator $\mathcal{T}e^{- i \int_{t_0}^t \hat{H}(s) ds}$ on the starting state $|\Psi(t_0)\rangle$  to get the state $|\Psi(t)\rangle$ at time $t$ using t-DMRG:
\begin{equation}
 |\Psi(t)\rangle=\mathcal{T}e^{- i \int_{t_0}^t \hat{H}(s) ds} |\Psi(t_0)\rangle.
\end{equation}
Here $\mathcal{T}$ is the time-ordering operator, and we set $\hbar=1$ in this paper.

More specifically, we first divide the time interval $t$ into a series of tiny time steps of the length $\tau$. The Hamiltonian is a function of time, but in each tiny time step it can be approximated by a constant, so we have
\begin{equation}
\mathcal{T}e^{- i \int_{t_0}^t \hat{H}(s) ds}\simeq e^{-i \tau \hat{H}(t-\frac{\tau}{2})} \cdots e^{-i \tau \hat{H}(\frac{3}{2}\tau)} e^{-i \tau \hat{H}(\frac{\tau}{2})}.
\end{equation}

We chose the the value of Hamiltonian in the middle of each interval to represent the Hamiltonian of that interval. At every time step we decompose $e^{- i \hat{H}(s) \tau}$ into local operators using second order Suzuki-Trotter decomposition, and we get
\begin{eqnarray}
 e^{- i \hat{H}(s) \tau}&=& e^{-i  \tau \left[\hat{H}_{d,0}(s)+\hat{H}_{0,1}+\hat{H}_{1,2}+\cdots+\hat{H}_{L-1,L}\right]}\nonumber \\ 
		    &=& e^{-i  \frac{\tau}{2} \hat{H}_{d,0}(s)}e^{-i  \frac{\tau}{2} \hat{H}_{0,1} }e^{-i  \frac{\tau}{2} \hat{H}_{1,2}}\nonumber \\
&& \cdots e^{-i \frac{\tau}{2} \hat{H}_{L-1,L}}e^{-i \frac{\tau}{2} \hat{H}_{L-1,L}}\cdots \nonumber \\
&& e^{-i  \frac{\tau}{2} \hat{H}_{1,2}}e^{-i  \frac{\tau}{2} \hat{H}_{0,1}} e^{-i \frac{\tau}{2} \hat{H}_{d,0}}+O(\tau^3),
\end{eqnarray}
where
\begin{equation}
\hat{H}_{d,0}(s)=\epsilon_d(s) d^\dagger d+(\frac{2 \Gamma D}{\pi})^{\frac{1}{2}}(f^\dagger_0 d+d^\dagger f_0),
\end{equation} 
and $H_{n,n+1}$ is the hopping term involving site $n$ and $n+1$. The only time-dependent part of the Hamiltonian is the impurity, so we only need to update the Suzuki-Trotter term of the impurity and the first site of the Wilson chain $e^{-i \frac{\tau}{2} \hat{H}_{d,0}(s)}$ at every time step.

We can also easily extend this method to study finite temperature dynamics. Instead of using infinite and finite DMRG to find the starting state, we use finite-temperature DMRG \cite{ftdmrg} to get the starting state. Then, one can evolve this purified state using t-DMRG to simulate the real-time dynamics at finite temperature \cite{schollwock05}. In this paper, however, we only focus on the zero temperature and noninteracting case.

\section{Exact Method}
The Hamiltonian~(\ref{Hwillsion}) is of quadratic form, so we can write it as 
\begin{equation}
   \hat{H}(t)=(a^\dagger_0, a^\dagger_1,  \cdots, a^\dagger_{L-1})H(t)(a_0, a_1,  \cdots, a_{L-1})^T,
\end{equation}
where  $a_0 \equiv d$, $a_i \equiv f_{i-1}$. $H(t)$ is a $L \times L$ Hermitian matrix with $L$ being the length of the Wilson chain.

By diagonalizing $H(t_0)$ we get
\begin{equation}
 \hat{H}(t_0)=\sum_k E_k \tilde a^\dagger_k \tilde a_k.
\end{equation}
The $k$-th single particle state is
\begin{equation}
 |k\rangle=\tilde a^\dagger_k|0\rangle =\sum_i u_{ik} a^\dagger_i |0\rangle,
\end{equation}
where $u_{ik}$ are the eigenvectors of $H(t_0)$, in the sense that $\sum_j H(t_0)_{ij}u_{jk}=E_k u_{ik}$.

At $t_0$ the system is in its ground state, characterized by the single-particle distribution function
\begin{equation}
f(k) = \left\{ \begin{array}{ll}
0, & E_k>0\\
1, & E_k<0
\end{array} \right..
\end{equation}
The initial density matrix of the whole system is
\begin{equation}
 \hat{\rho}(t_0)=\sum_k f(k) |k\rangle\langle k|.
\label{dm_init}
\end{equation}
The density matrix evolves according to the von Neumann equation
\begin{equation}
 i \frac{\partial \hat{\rho}(t)}{\partial t}=[\hat{H}(t),\hat{\rho}(t)].
\end{equation}
This equation can easily be solved with an ordinary differential equation solver such as Matlab's ode45.
Then we can calculate the expectation value of operators, like $\hat{n}_d(t)$, as
\begin{equation}
 n_d(t) = {\rm Tr}\left[\hat{n}_d \hat{\rho}(t)\right]= {\rm Tr}\left[a^\dagger_0 a_0 \hat{\rho}(t)\right].
\end{equation}

\section{Results and Physical Interpretation}

\begin{figure}[htbp]
\centering
\begin{tabular}{l}
\includegraphics[width=0.43\textwidth]{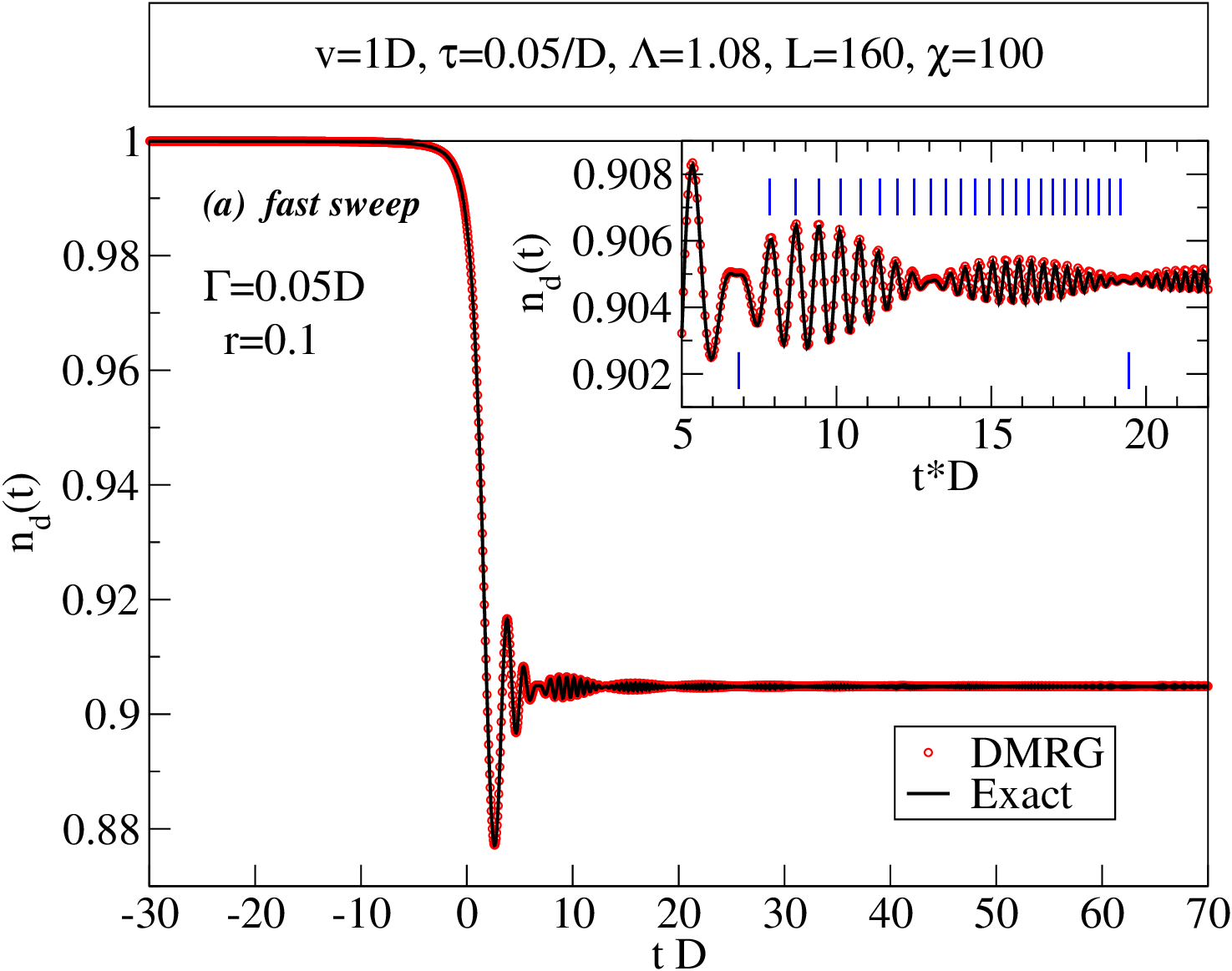} \\
\includegraphics[width=0.43\textwidth]{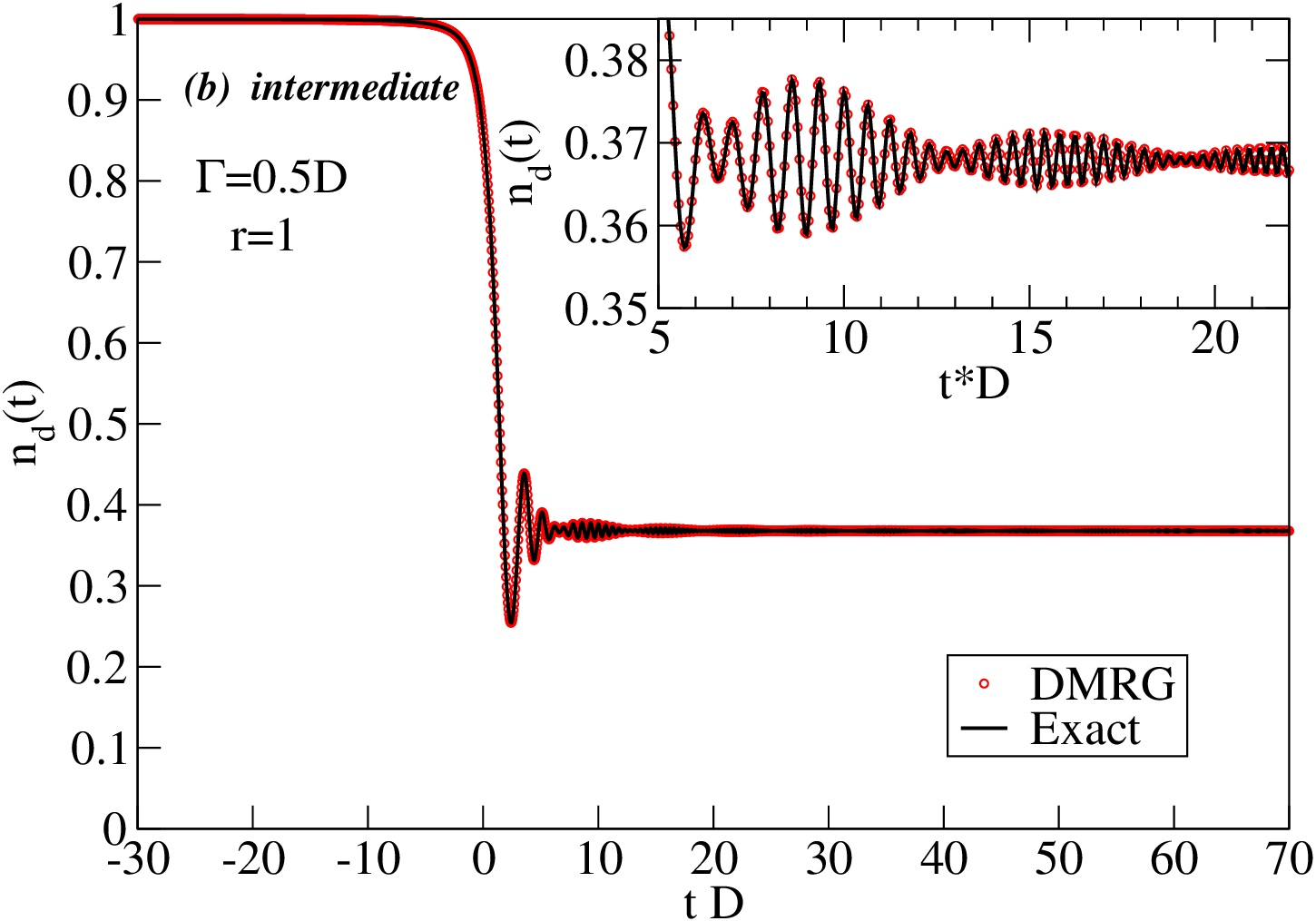} \\
\includegraphics[width=0.43\textwidth]{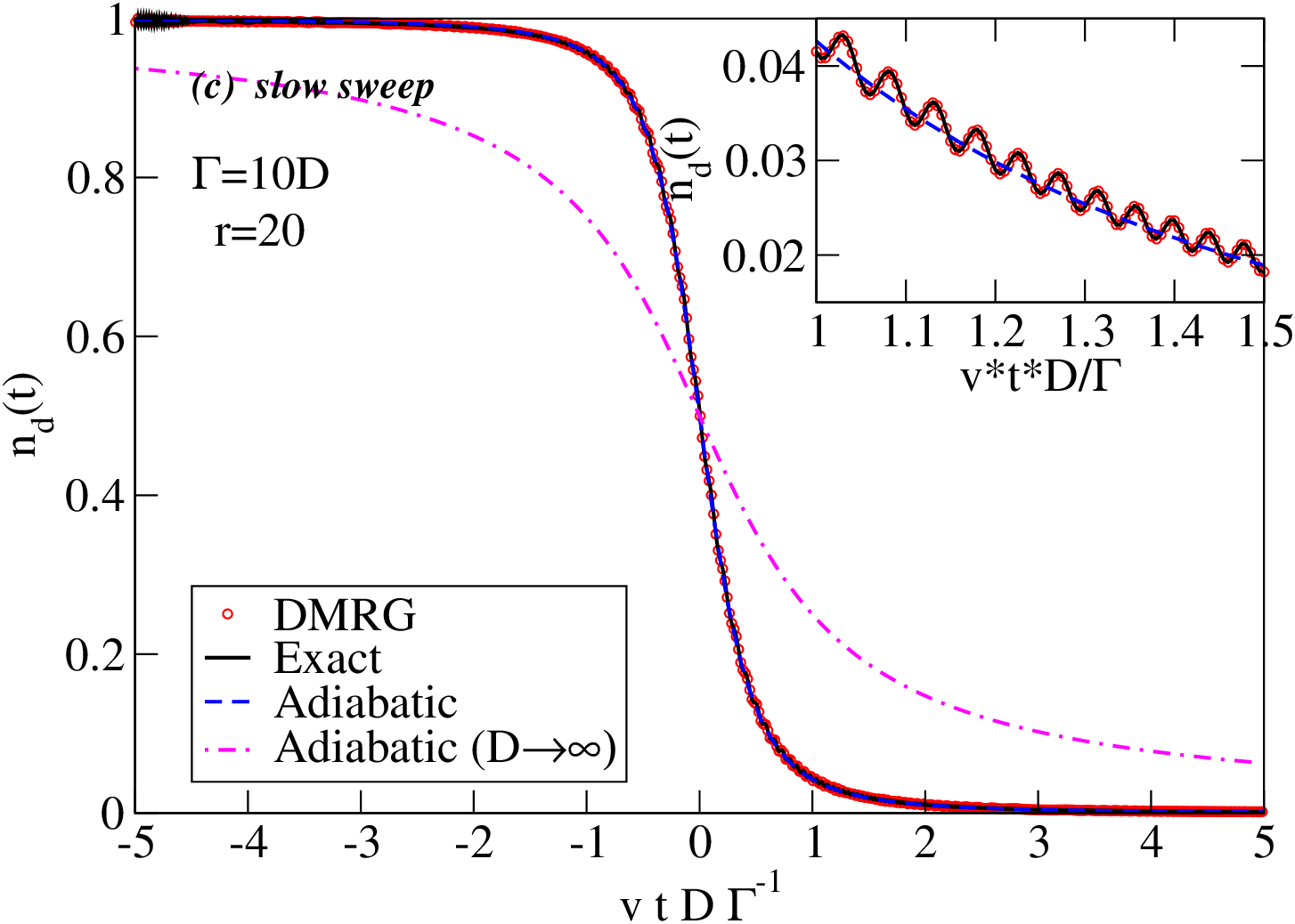}
\end{tabular}
\caption{(Color online) The local occupation number $n_d(t)$ as a function of time, calculated with both exact and DMRG method in the three parameter regimes. At the top, we give the choices made for the following parameters: sweeping speed $v$, Suzuki-Trotter step $\tau$, logarithmic discretization parameter $\Lambda$, Wilson-chain length $L$, and the number of states kept in DMRG calculation $\chi$. The value of hybridization parameter $\Gamma$ and the corresponding dimensionless parameter $r \equiv 2 \Gamma /v$ are given in each figure respectively. The insets zoom in on fine details of the curves. (a) The markers in the inset indicate the periods of the oscillations and beats obtained from the simple physical picture discussed in the text [see Eq.~(\ref{beats})]. (c) The dashed lines are the reference results of the adiabatic sweep calculation, and the dash-dotted line is the adiabatic sweep result in infinite band limit, which is simply $n_d(\frac{\epsilon_d}{\Gamma})=\frac{1}{2}-\frac{1}{\pi}\arctan{\frac{\epsilon_d}{\Gamma}}$.}
\label{3regimes}
\end{figure}

In Fig.~\ref{3regimes} we plot both the exact and DMRG results in the three typical parameter regimes at zero temperature respectively. We use Wilson-chain length $L=160$ and logarithmic discretization parameter $\Lambda=1.08$ for all the three figures. We will discuss the discretization method in more detail in the next section. Note that we set $D=1$ in our calculation.

\begin{figure}[htbp]
\centering
	\includegraphics[width=0.45\textwidth]{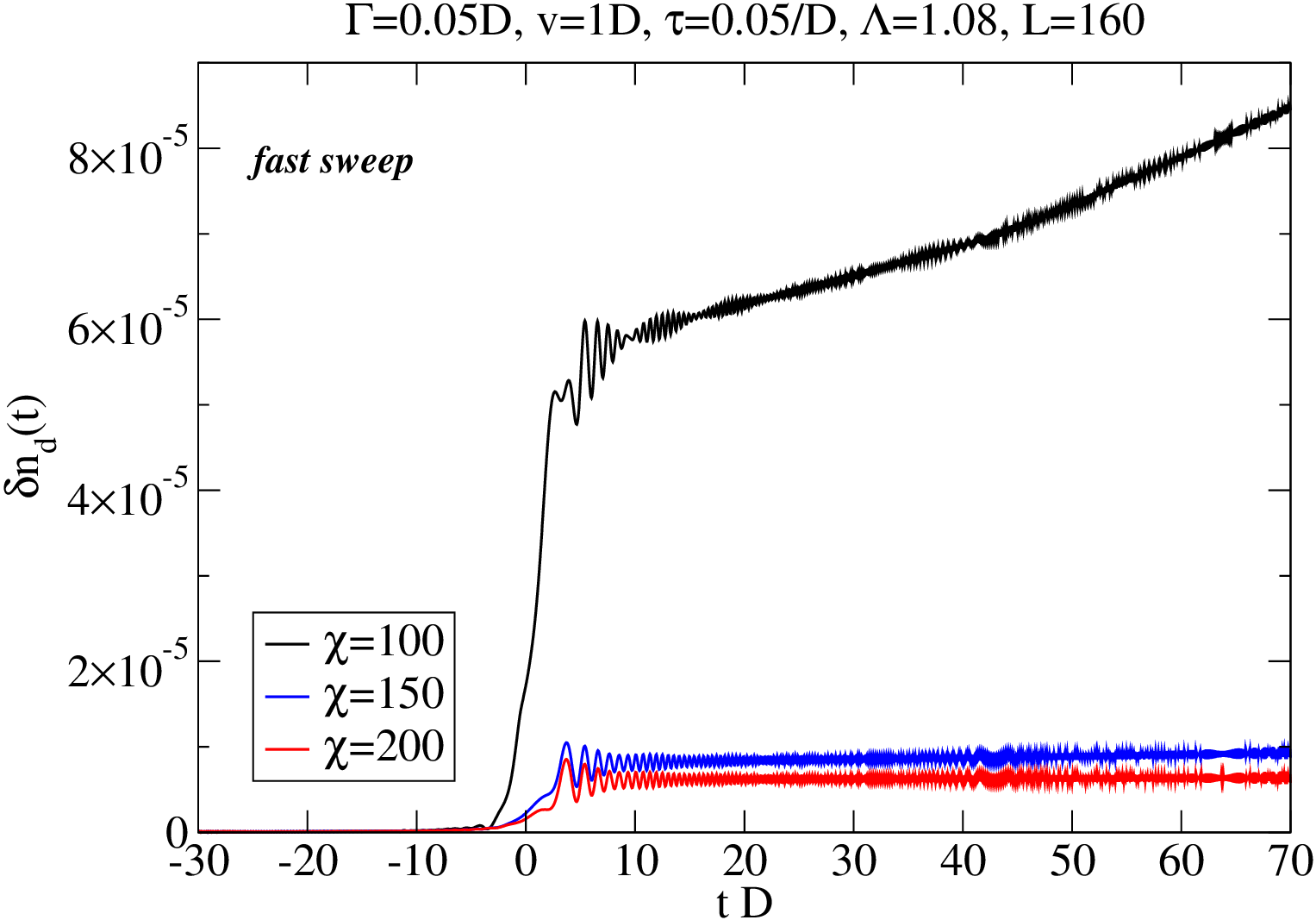}
	\caption{(Color online) Error of the DMRG results for $\delta n_d(t) \equiv n_d^{\rm DMRG}(t)-n_d^{\rm Exact}(t)$ in fast sweep regime when keeping 100, 150 and 200 states.}
	\label{error}
\end{figure}

For all three regimes, the DMRG error (shown in Fig. \ref{error} for fast regime) is at worst of order $10^{-4}$ when keeping $\chi=100$ states during DMRG calculation. This error can be further reduced by increasing $\chi$.

Let us now try to understand the results physically. In the fast sweep regime  the spinless fermion on the impurity does not have enough time to totally jump into the band, so the occupation on the impurity $n_d(t)$ converges to a finite value as the level is swept through and out of the band. In contrast, in the slow sweep regime the fermion ends up in the band with a very high probability. For comparison we also show the results of an adiabatic sweep in the slow sweep regime in Fig.~\ref{3regimes}. The adiabatic results are obtained from the thermodynamic average ${\rm Tr}[\hat{\rho}_{\epsilon_d(t)} \hat{n}_d]$, where $\hat{\rho}_{\epsilon_d(t)}$ is calculated using Eq.~(\ref{dm_init}) with single particle states $|k_{\epsilon_d(t)}\rangle$ of the Hamiltonian $H_{\epsilon_d(t)}$. Evidently, the DMRG and exact results agree very well with the adiabatic results.

Another important feature of the results is the oscillation of $n_d(t)$. To understand it, we first study a simplified model, in which we only consider one level in the band and disregard the rest levels for the moment. When there is one spinless fermion in this system the Hamiltonian is
\begin{equation}
H(t)=\left(\begin{array}{cc}
E_{0}(t) & \gamma \\
\gamma & E_{1}\end{array}\right),
\label{Hsimple}
\end{equation}
This is just the Hamiltonian of the original Landau-Zener problem. We denote the instantaneous two eigenstates as $|+\rangle_{t}, |-\rangle_{t}$ with the corresponding eigenenergies $E_{\pm}(t)=\frac{1}{2}[E_0(t)+E_1 \pm \omega(t)]$, where 
\begin{equation}
 \omega(t)=\sqrt{4\gamma^2+(E_1-E_0(t))^2}.
\end{equation}

The probability that a state of the form $|\phi(t)\rangle=a|-\rangle_{t}+b|+\rangle_{t}$ at time $t$ will still be found in the same state at time $t+\delta t$, is given by
\begin{subequations}
\label{freq}
\begin{eqnarray}
\label{freq_a}
\tilde{P}(t)&\equiv&|\langle\phi(t)|\phi(t+\delta t)\rangle|^{2}  \\
\label{freq_b}
    &=& |a|^4+|b|^4+2|a b|^2\cos\left[ \omega(t) \delta t \right].
\end{eqnarray}
\end{subequations}

In each time interval, the instantaneous oscillation frequency $\omega(t)$ of $\tilde{P}(t)$ is equal to the instantaneous oscillation frequency of $|\langle\phi(t_0)|\phi(t)\rangle|^{2}$ to the zeroth order in $\delta t$. Therefore, the probability for the system initially in a state $|\phi(t_{0})\rangle$ to still be found in this state at a later time $t$, 
\begin{equation}
P(t)\equiv|\langle\phi(t_{0})|\phi(t)\rangle|^{2},
\end{equation}
will have an oscillating component proportional to $\cos\left[ \int_{t_0}^t \omega(s) d s\right]$.

We now return to the original problem and use the picture described above to roughly estimate the period of the oscillations in the fast sweep regime. In the fast sweep regime according to Pauli exclusion principle the influence of the unoccupied levels of the upper half of the band is dominant. We can neglect the lower half of the band, and add up the contributions of all levels $E_1$ in the upper half band to the oscillations by integrating the above mentioned cosine term over the energies $E_1$. Therefore the occupation on the resonant-level 
\begin{equation}
n_d(t) \approx \int^D_0 P(t) dE_1
\end{equation}
will contain an oscillating contribution proportional to 
\begin{equation}
\sin\left[\frac{D}{2}(t-t_0) \right] \cos \left[\frac{1}{2} (v t^2-D t+D t_0-v t_0^2)\right].
\label{beats}
\end{equation}
To get the above result, we approximated $\omega(t)$ by $E_1-v t$, neglecting the term $4\gamma^2=4\Gamma D/\pi$ ($\gamma \equiv \sqrt{D \Gamma /\pi}$ is the prefactor of the hybridization term in the energy representation of Hamiltonian~(\ref{Horiginal})\cite{nrg80}.). This is a good approximation, except around $t=0$, when the local level is near the middle of the band, and $|E_0 - E_1|$ is not significantly larger than $\Gamma$. 

The resulting Eq.~(\ref{beats}) can be used to understand the nature of the oscillations and beats observed in the fast sweep regime in Fig.~\ref{3regimes}. The factor $\sin\left[D(t-t_0)/2 \right]$ is the beat, and the period of the beats is $T_{\rm beats} =4 \pi/D$. We plot two markers with a separation of $4\pi/D$ under the curve in the inset of Fig.~\ref{3regimes}(a); they fit the period of the beats very well. The markers above the curve in the insets of Fig.~\ref{3regimes}(a) are obtained by solving 
\begin{equation}
 \frac{1}{2} (v t^2-D t+D t_0-v t_0^2)=2 m \pi+ \rm const,
\label{osc}
\end{equation}
where $m$ is an integer such that the markers are best aligned with the maxima of the oscillations shown. We can see that the final agreement in position is excellent.

\begin{figure}[htbp]
\centering
	\includegraphics[width=0.43\textwidth]{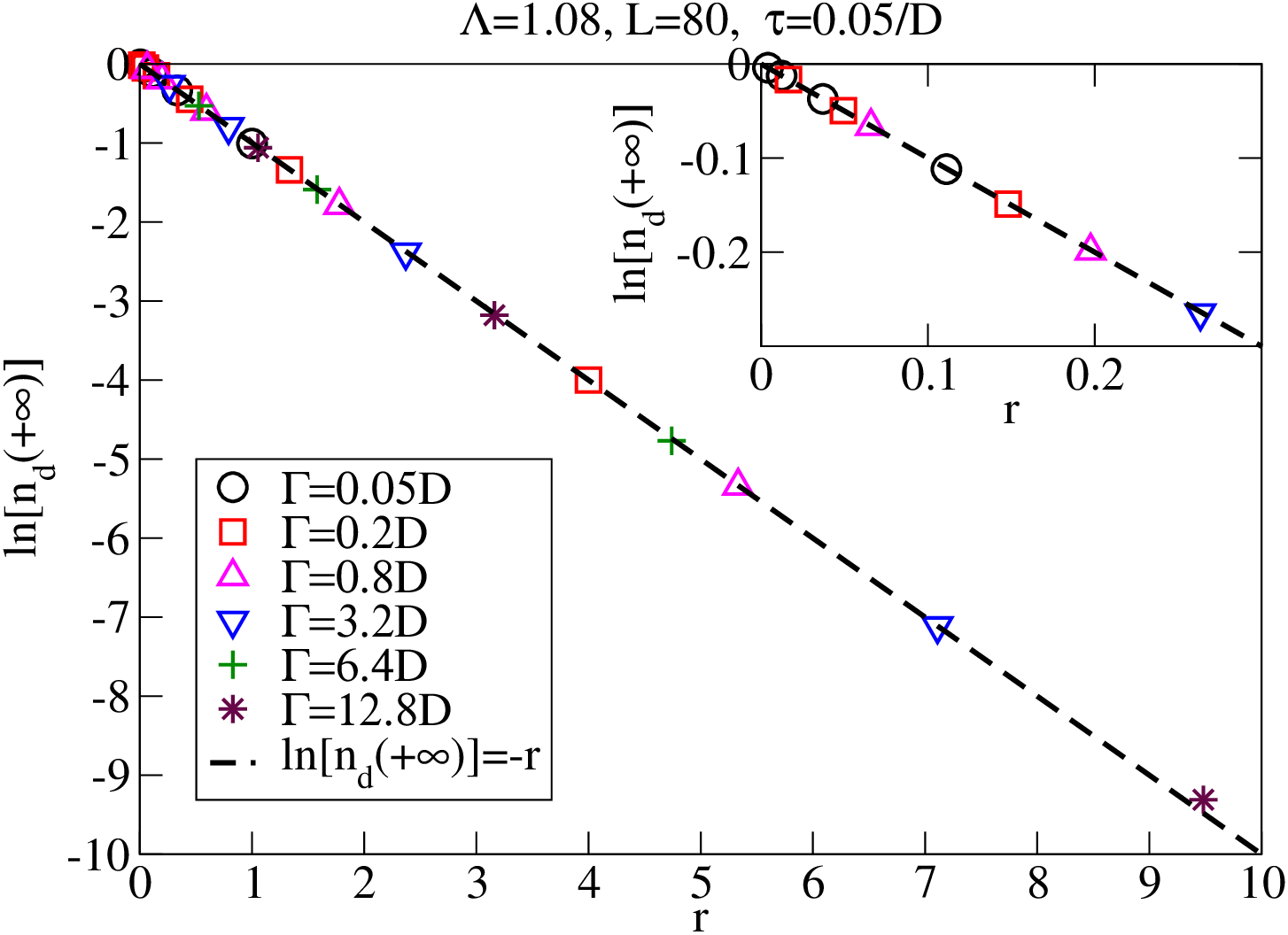}
	\caption{(Color online) Exact results checking the relation between the final local occupation number $n_d(+\infty)$ and $r$. Numerically, we approximate $n_d(+\infty)$ by averaging $n_d(t)$ of the last 4 time steps. The time span we use here is $t \in [-200/D, 200/D]$.  To get $n_d(+\infty)$ at different $r$, we choose 6 different $\Gamma$ from a wide parameter regime, and with each $\Gamma$ 6 different sweeping speed: $v=0.1D, 0.3D, 0.9D, 2.7D, 8.1D, 24.3D$ are used to calculate $n_d(+\infty)$. We only plot the data for $r<10$ because the accumulated numerical error becomes significant compared to $n_d(+\infty)$ for $r>10$. The dashed line is a reference line of $\ln[n_d(+\infty)]=-r$. The inset zooms in on small $r$.}
	\label{fnd}
\end{figure}

Last but not the least, we examined the dependence of the final local level occupation number $n_d(+\infty)$ on $r$ (shown in Fig.~\ref{fnd}), and find it has the typical Landau-Zener exponential relation:
\begin{equation}
 n_d(+\infty)=e^{-r}.
\end{equation}

\begin{table}[htdp]
\caption{The $n_d(+\infty)$ data used in Fig.~\ref{fnd}}
\begin{center}
\begin{ruledtabular}
\begin{tabular}{c| c c c c c c}
\backslashbox{$\Gamma$}{$v$}	&	0.1D		&	0.3D		&	0.9D		&	2.7D		&	8.1D		&		24.3D\\
\hline
0.05D						&	0.36516	&	0.71499	&	0.98419	&	0.96345	&	0.98766	&	0.99587\\
%\hline
0.2D							&	0.01831	&	0.26199	&	0.63925	&	0.86164	&	0.95155	&	0.98358\\
%\hline
0.8D							&	-		&	0.00480	&	0.16758	&	0.55114	&	0.81984	&	0.93593\\
%\hline
3.2D							&	-		&	-		&	0.00081	&	0.09221	&	0.45177	&	0.76732\\
%\hline
6.4D							&	-		&	-		&	-		&	0.00850	&	0.20404	&	0.58877\\
%\hline
12.8D						&	-		&	-		&	-		&	0.00009	&	0.04162	&	0.34660\\
\end{tabular}
\end{ruledtabular}
\end{center}
\label{default}
\end{table}%

This agrees with previous analytical results\cite{sinitsyn_multiparticle_2002,Wubs06}. Note that though $n_d(+\infty)$ only depends on $r$, the detailed structure of the $n_d(t)$ curve is determined by $v$ and $\Gamma$ respectively. (See Eq.~(\ref{osc}) for example.)

\section{Role of Discretization Parameter}

\begin{figure}[htbp]
\begin{tabular}{l}
\label{difflambda}
\includegraphics[width=0.435\textwidth]{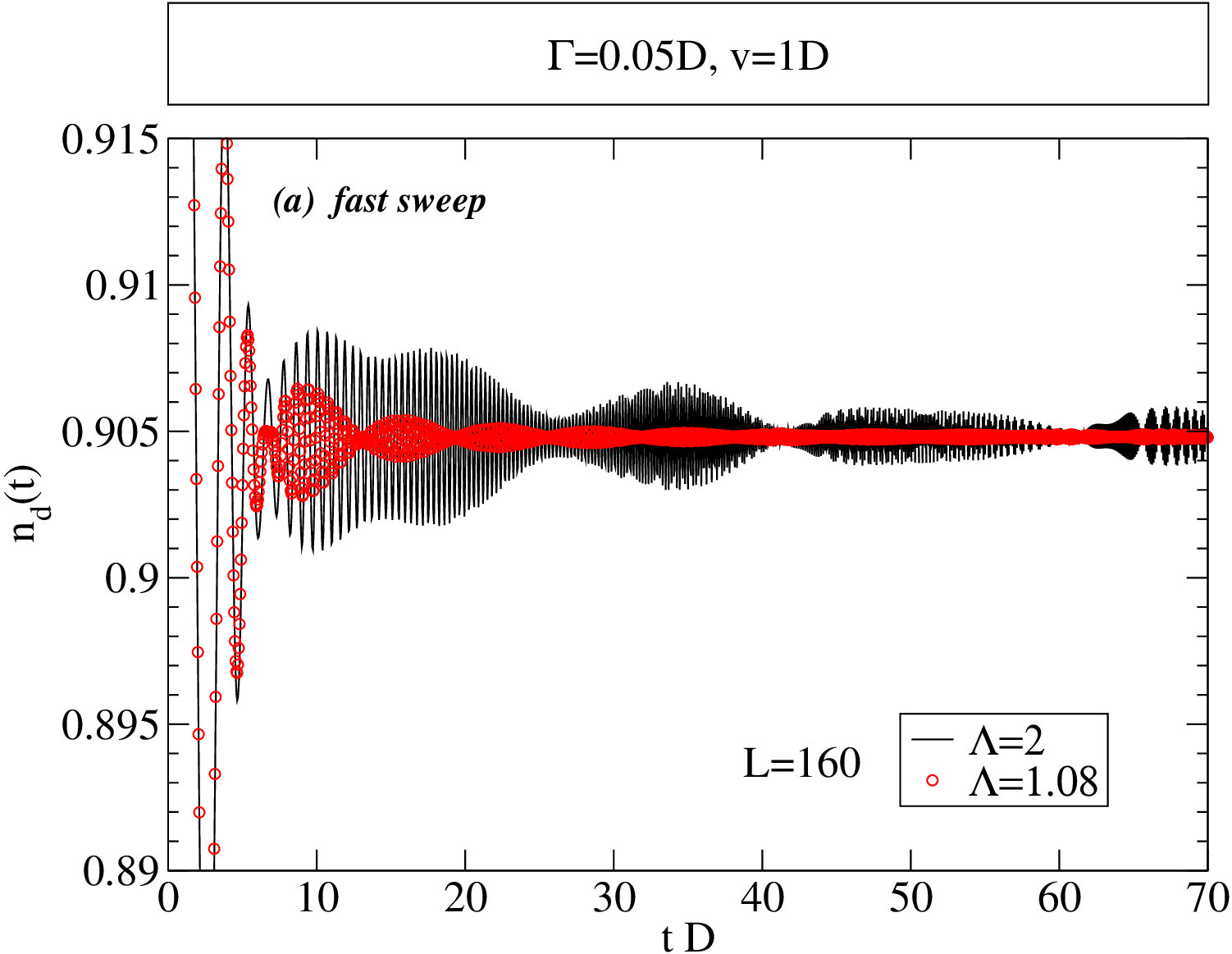} \\
\label{linlog}
\includegraphics[width=0.435\textwidth]{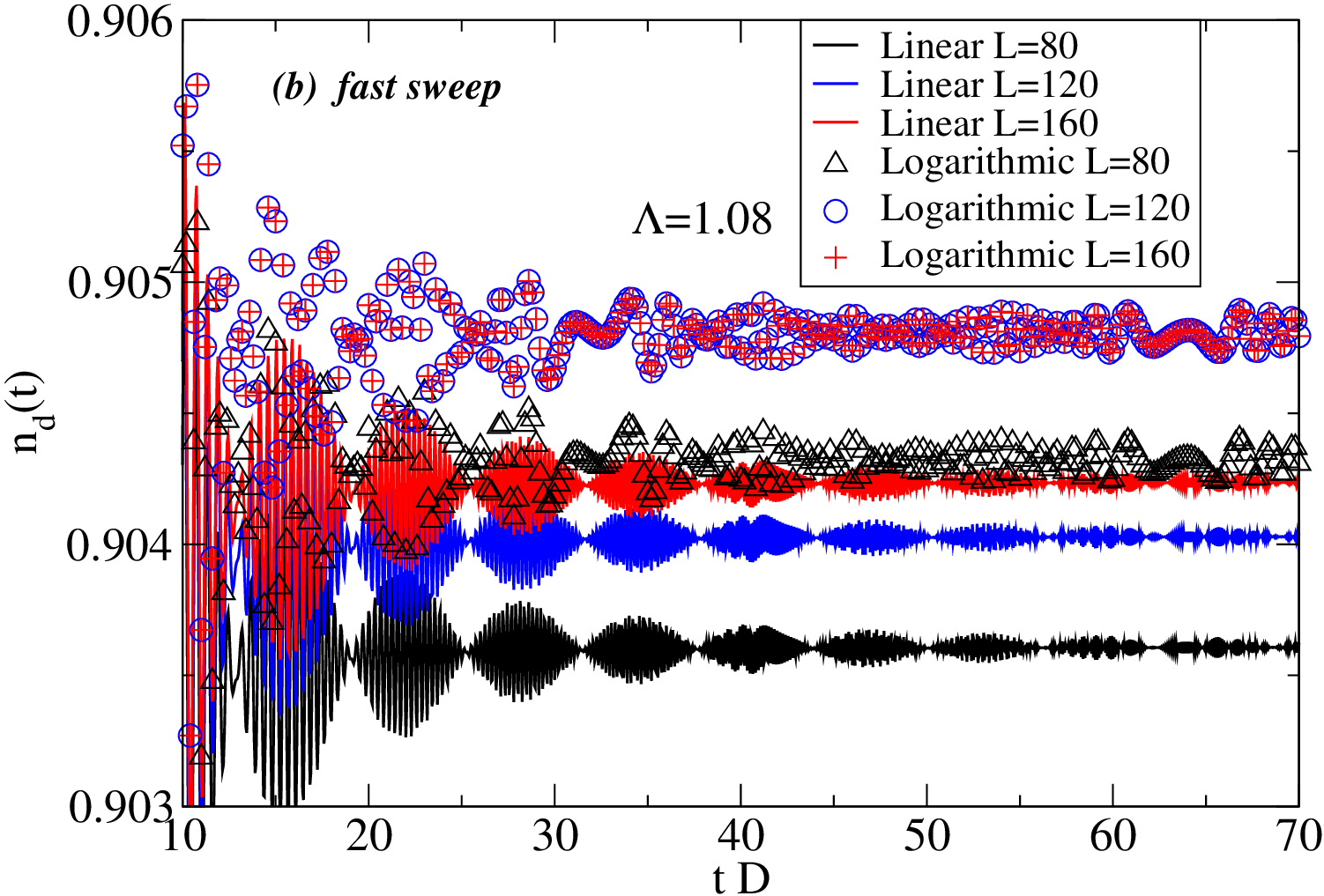} \\
\label{finalp_linlog}
\includegraphics[width=0.435\textwidth]{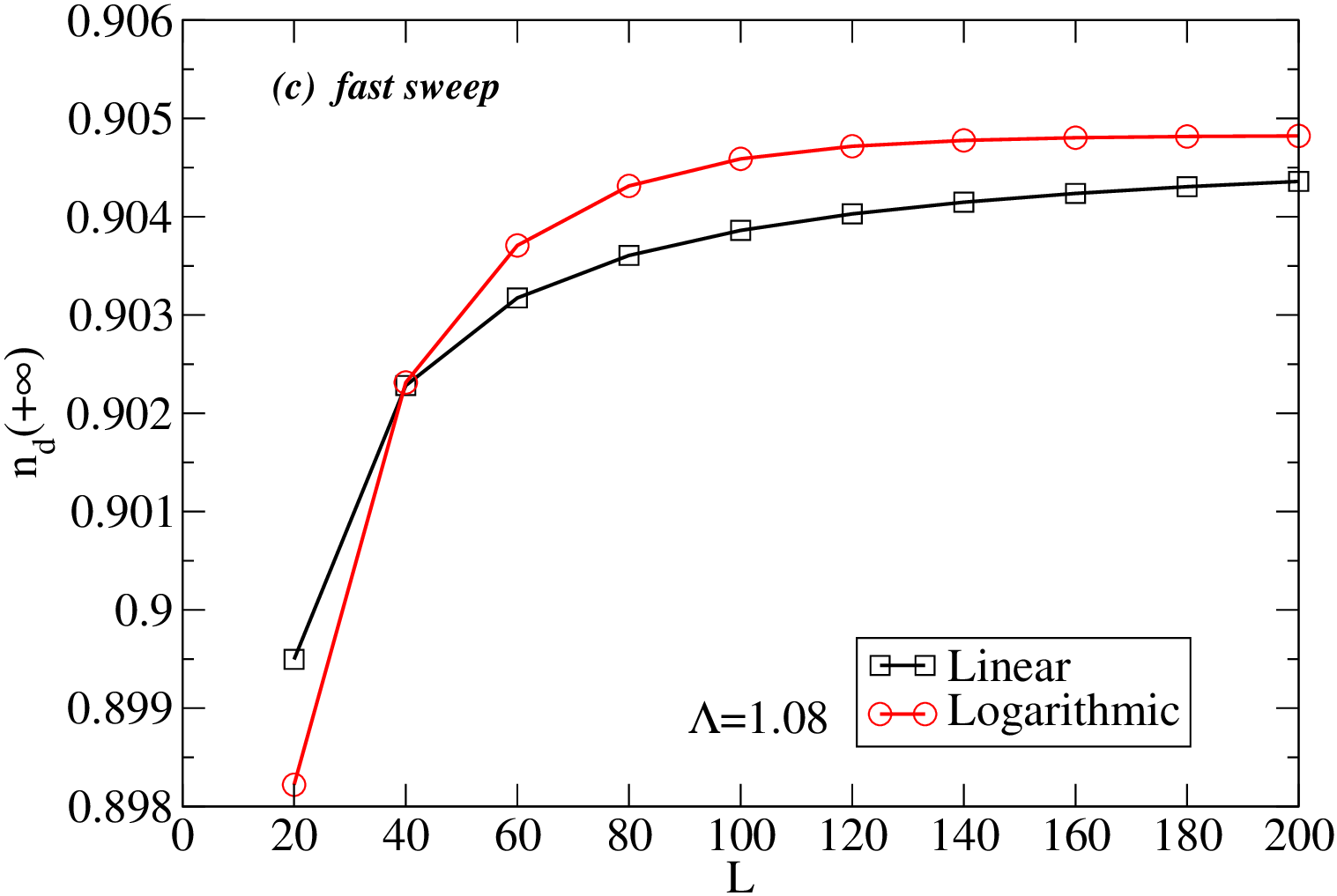}
\end{tabular}
\caption{(Color online) Comparison of the exact results of different logarithmic discretization parameters. Both figures zoom in on fine details. Here we study fast sweep regime as an example. (a) Comparison of the results with different $\Lambda$. The large oscillations for $\Lambda=2$ for later times are due to the relatively coarse discretization and hence artificial. (b),(c) Comparison of the converging speed with respect to the Wilson-chain length $L$ of linear and logarithmic discretization method. }
\label{discretization}
\end{figure}

As in NRG, the value chosen for the discretization parameter can affect the real-time dynamics, if it does not lie sufficiently close to 1. Fig.~\ref{discretization}(a) compares the exact results of $\Lambda=1.08$ and $\Lambda=2$ in fast sweep regime. Note that for $\Lambda=2$, big oscillations in $n_d(t)$ remain long after the transition. These are artificial consequences of the rather coarse discretization scheme, which diminish strongly as $\Lambda$ is reduced towards 1. Indeed, for $\Lambda=1.08$, most of these oscillations have disappeared. Further reduction of $\Lambda$ does not change the results significantly anymore. Note that, incidentally, the ability of allowing a logarithmic discretization parameter very close to 1 is a big advantage of DMRG over NRG.

With the physical picture described in the last section, we can also understand why there are artificial oscillations if $\Lambda$ is big. If we use a big logarithmic discretization parameter, the part of the band far away from the Fermi level is poorly represented by only a few levels, which means that the oscillations from different levels do not average out as well as would have been the case for a true continuum of levels.

We use logarithmic discretization instead of linear discretization because in the problem we studied, the levels near Fermi surface contribute more than levels far away from it, and logarithmic discretization represents the part of band around Fermi surface more efficiently.\footnote{A systematic way of optimizing the discretization scheme, base on analyzing the contribution of each level from the discretized band to the reduced density matrix of the local level, was recently proposed by Zwolak\cite{zwolak_finite_2008}.} This is reflected in the convergence of the results with respect to the Wilson-chain length $L$ shown in Fig.~\ref{discretization}. As other parameters are the same, the two discretization methods will both converge to the same result when $L \to \infty$. Therefore the faster the result converges the better the method is. We can see from Fig.~\ref{discretization} (b) that the difference of $n_d(t)$ between $L=120$ and $L=160$ chains is already negligible for the case of logarithmic discretization while still significant if using linear discretization, which means the results converge more quickly if we use logarithmic discretization. This is even more obvious by comparing the convergence speed of $n_d(+ \infty)$ shown in Fig.~\ref{discretization}(c).

\section{Conclusions and Outlook}

By studying a benchmark model we demonstrated that the t-DMRG is a very accurate method to calculate real-time dynamics of quantum impurity system with a time-dependent Hamiltonian. To compare with the exact results, the model we studied here is a non-interacting model, but DMRG can also treat interacting problems similarly. 

Though t-DMRG cannot calculate arbitrary long times (in contrast to TD-NRG) it can give reliable results in a relatively long time which we expect to be long enough for numerous practical purposes. For example, in quantum information, where fast quantum processes are more useful, the relevant physics happens in a relatively short time scale, which can be simulated by t-DMRG with a high precision.  We thus expect t-DMRG to be a powerful tool to study the real-time dynamics of quantum impurity systems, in particular in the context of modeling the dynamics of damped, driven qubits.

\begin{acknowledgments}
We gratefully acknowledge fruitful discussions with Theresa Hecht and Wolfgang M\"under, and Barbara Englert for help in editing the text. We also would like to thank Peng Zhang, Shaojing Qin, Gang Yang and Qiaoni Chen for helpful discussions. This work was supported by the DFG (SFB 631, SFB-TR12, De-730/3-2). Financial support of the German Excellence Initiative via the Nanosystems Initiative Munich (NIM) is gratefully acknowledged.
\end{acknowledgments}

\end{document}